\documentclass[twocolumn]{aastex63}        
\usepackage{graphicx,times}

\received{2020 January 27}
\revised{2020 March 30}
\accepted{2020 April 3}
\submitjournal{ApJL}

\shorttitle{Optical counterpart of UGC\,6456~ULX}
\shortauthors{Vinokurov et al.}

\begin{document}
\title{Optical Counterpart to the Ultraluminous X-Ray Source in the UGC\,6456 Galaxy}

\correspondingauthor{A. Vinokurov}
\email{vinokurov@sao.ru}

\author[0000-0001-5197-2457]{A. Vinokurov}
\affiliation{Special Astrophysical Observatory, Nizhnij Arkhyz, 369167, Russia}
\author[0000-0002-8816-2017]{K. Atapin}
\affiliation{Sternberg Astronomical Institute, Lomonosov Moscow State University, Universitetskij Pr. 13, Moscow 119992, Russia}
\author{Y. Solovyeva}
\affiliation{Special Astrophysical Observatory, Nizhnij Arkhyz, 369167, Russia}

\begin{abstract}
We report the identification of the optical counterpart to the transient ultraluminuos X-ray source in the blue dwarf galaxy UGC\,6456 (VII\,Zw\,403). The source is highly variable in both the X-ray (more 100 times, 0.3--10~keV) and optical (more 3 times, V band) ranges. The peak X-ray luminosity of UGC\,6456~ULX exceeds $10^{40}$~erg\,s$^{-1}$; the absolute magnitude when the source is optically bright is M$_V = -8.24\pm0.11$, which makes this source one of the brightest ULXs in the optical range. We found a correlation between the optical and X-ray fluxes (with a coefficient of $0.9\pm 0.3$), which may indicate that the optical emission is produced by re-processing of the X-rays in outer parts of the optically-thick wind coming from the supercritical accretion disk. Optical spectra of UGC\,6456~ULX show broad and variable hydrogen and helium emission lines, which also confirms the presence of the strong wind. 
\end{abstract}
\keywords{accretion, accretion disks --- X-rays: binaries --- X-rays: individual: UGC\,6456~ULX }

%

%
\section{Introduction}           
\label{sect:intro}
Ultraluminous X-ray sources (ULXs) are variable objects whose luminosity exceeds the Eddington limit for stellar mass black holes ($\geq 10^{39}$ erg~s$^{-1}$), assuming isotropic emission. These objects are located outside the centers of galaxies, that is, they are not super-massive black holes. Early papers on ULXs suggested intermediate mass black holes (IMBHs; $10^2-10^4$ M$_{\odot}$) as accretors in these systems (e.g., \citealt{Colbert1999ULXIMBH}). However, studies of the last decade \citep{Gladstone2009,Sutton2013,Fabrika2015,Pinto2016Nat,Walton2018puls} have shown that observational properties of most ULXs can be explained by supercritical gas accretion onto black holes of stellar masses (from a few to several tens masses of the Sun), or even onto neutron stars. The last possibility has been convincingly confirmed by detection of coherent X-ray pulsations in six ULXs \citep{Bachetti2014Nat,Furst2016pulsP13,Israel2016pulsNGC5907,Carpano2018NGC300,Sathyaprakash2019pulsNGC1313X2,RodriguezCastillo2019pulsM51ulx7}. 

In contrast to the X-ray range, in the optical one ULXs are studied much more poorly. Only $\gtrsim 20$ ULXs have been unequivocally identified using the {\it Hubble Space Telescope} ({\it HST}) data. All of them are faint sources with visual magnitudes of $m_V = 21 - 26$ \citep{Tao2011counterparts,Gladstone2013counterparts}. Moreover, most ULXs are associated with star-forming regions (e.g., \citealt{Poutanen2013AntennaeULX}) and located in crowded regions, which makes it difficult to study them with ground-based telescopes.

Here we present the identification of the optical counterpart to the ULX in the galaxy UGC\,6456 (VII~Zw~403), which is one of the closest blue compact dwarf galaxies (D = 4.54 Mpc, \citealt{Tully2013}). UGC\,6456~ULX is a transient ULX: its X-ray luminosity changes by more than two orders of magnitude with a peak value of $1.7 \times 10^{40}$ erg\,s$^{-1}$ in the 0.3--8~keV energy range \citep{Brorby2015UGC6456}. In the bright state, the sources shows a very hard power-law spectrum with a photon index of $\Gamma \sim 1$. We report the presence of a correlation between the long-term X-ray and optical variability of UGC\,6456~ULX and present results of the optical spectroscopy.

\begin{figure*} 
\centering
\includegraphics[angle=0,scale=0.358]{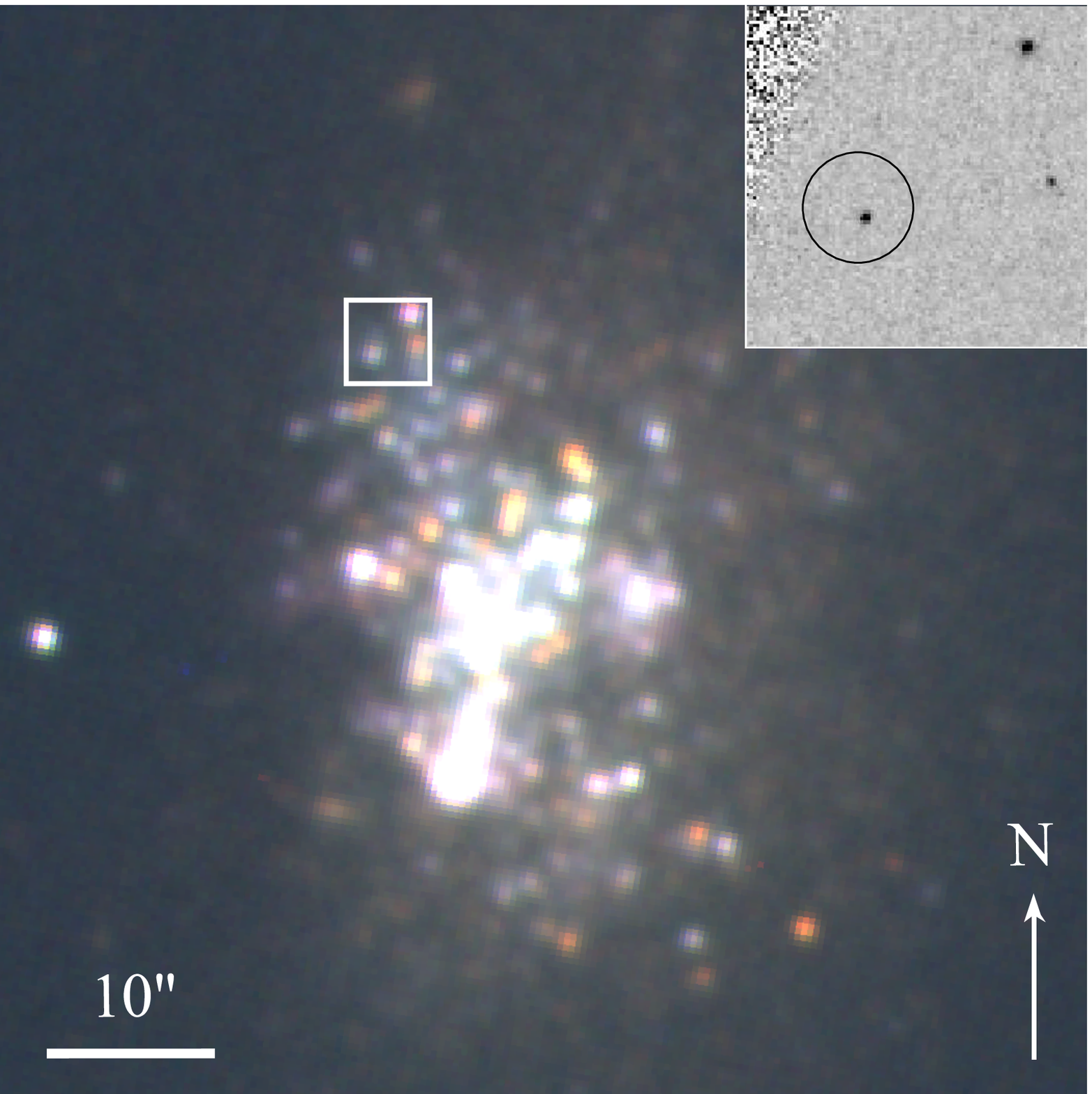} 
\hspace{0.25cm}
\includegraphics[angle=0,scale=0.49]{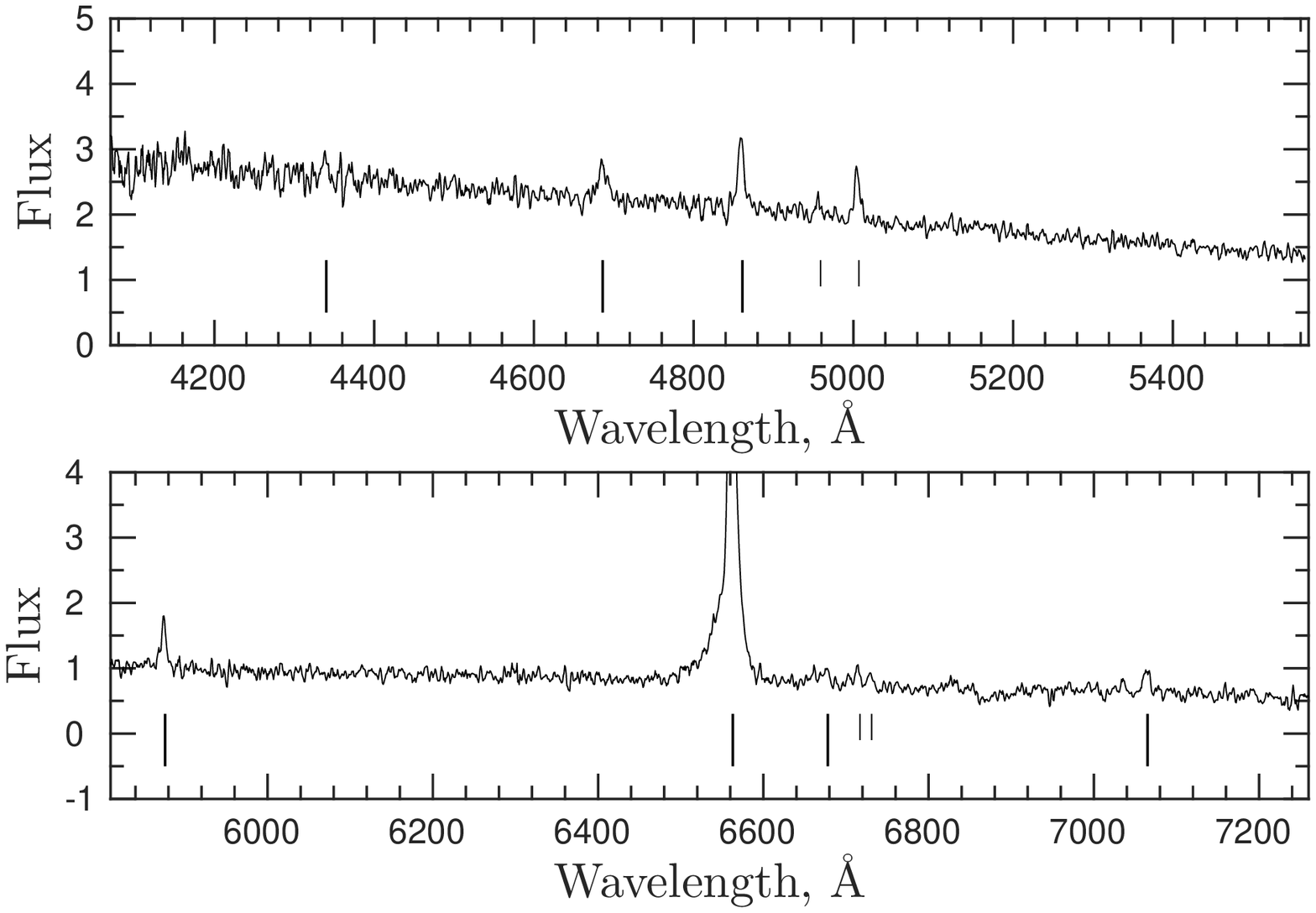}  
\caption{{\it Left panel}: superposition of B, V and R$_c$ images of the UGC\,6456 galaxy taken with BTA/SCORPIO. The inset shows the HST/WFPC2/F555W image of the region around UGC\,6456~ULX marked by the square; the circle indicate the 0.8$\arcsec$ error box of the ULX position derived from the {\it Chandra} data. {\it Right panel}: blue (top) and red (bottom) spectra of the UGC\,6456~ULX optical counterpart with the best signal-to-noise ratio. The spectra are not simultaneous. Broad emission lines are designated by long vertical bars; the narrow emissions produced by the nearest nebula are marked by short bars (see text for details).} 
\label{Fig1}
\end{figure*}

\section{Astrometry}

To identify the optical counterpart of UGC\,6456~ULX, we used archival images from the {\it Chandra X-Ray Observatory} and {\it HST}. {\it Chandra} observed the source only once, on 2000 January 7 (ObsID\,871). The source was positioned on the S3 chip of the Advanced CCD Imaging Spectrometer (ACIS) with a very small offset from the optical axis ($<0\farcm1$). The ULX coordinates in the ACIS image obtained with the Wavdetect task of CIAO\,4.11 is R.A.(J2000)$\,=\,11^{\mathrm h}\,28^{\mathrm m}\,03\fs000$, Decl.(J2000)$\,=\,+78^\circ\,59\arcmin\,53\farcs41$ with a statistic error of $\sim0\farcs1$ and an absolute astrometric uncertainty of $0\farcs8$ at 90\% confidence level\footnote{http://cxc.harvard.edu/cal/ASPECT/celmon/}.

In the optical range, we have chosen the HST observation taken on 1994 February 16 with the Wide Field and Planetary Camera~2 (WFPC2) in the F555W filter. To apply astrometric corrections to the HST image, we used four reference stars that present in the Gaia Data Release~2 \citep{Gaia2018}. After accounting for the shifts, derived standard deviation of the difference between Gaia and corrected {\it HST} positions of the reference stars became less than $0\farcs03$, and the resultant absolute astrometric uncertainty about $0\farcs02$ at the 90\% confidence level.

There is only one relatively bright object in the WFPC2/F555W image within the {\it Chandra} $0\farcs8$ error circle of UGC\,6456~ULX (left panel of Figure \ref{Fig1}). It is a point-like source with a visual magnitude of $m_{\mbox{\scriptsize F555W}} = 21.59\pm0.06$ ($m_V = 21.58$) in the Vegamag system\footnote{Photometry was performed on \texttt{c0f} image in HSTPHOT\,1.1 \citep{Dolphin2000HSTWFPC2}}. At the galaxy distance this corresponds to an absolute magnitude of M$_V = -6.9$ after correction for reddening of A$_V = 0.2$ mag (see Section \ref{sect:spec}).

\section{X-ray and optical variability}
\label{sect:XrayOptica}

\begin{figure*}
\centering 
\includegraphics[angle=0,scale=0.8]{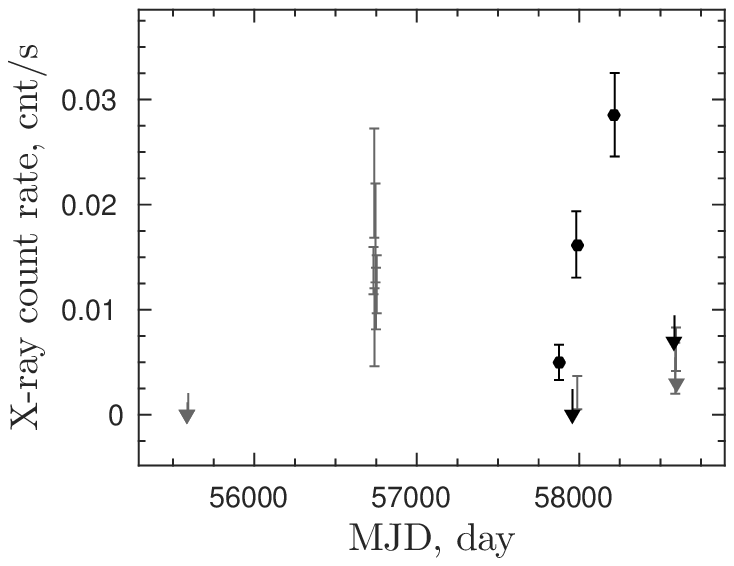}
\hspace{0.5cm}
\includegraphics[angle=0,scale=0.8]{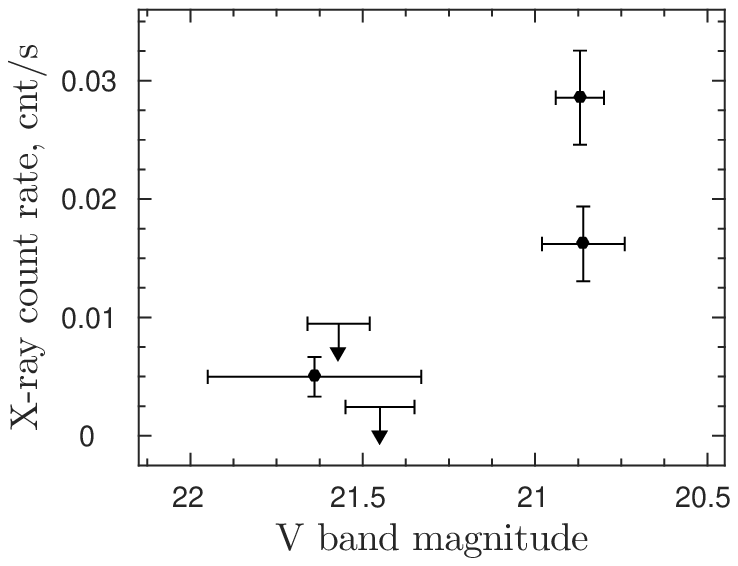} 
\caption{{\it Left panel}: net count of UGC\,6456~ULX in the 0.3--10~keV range measured from the {\it Swift} data. The data points shown by black symbols indicate those Swift observations that were carried out quasi-simultaneously with the optical ones (Table~\ref{Tab1}). The arrows show an 90\%\ upper limit to the count rate for the observations that have accumulated less than 4 counts. {\it Right panel}: net count rate plotted against V-band magnitude obtained from the quasi-simultaneous observations.}
\label{Fig2} 
\end{figure*} 

The ULX was observed 17 times with the X-ray telescope of the Neil Gehrels Swift Observatory ({\it Swift}/XRT) since January 2011. From 2014 March 16 through 2014 April 6 the observations were made every few days. Exposure times were in the range of 230--4170\,s. To extract a light curve we utilized preprocessed event files provided by the UK Swift data centre\footnote{https://www.swift.ac.uk}. The source events were extracted from a circular region of 30 arcsec radius using the XSELECT task of \textsc{HEASOFT}\,6.26.1. The background was taken from an annulus with inner and outer radii of 80 and 130~arcsec, respectively, centered on the source. The resulting light curve is shown in Fig.~\ref{Fig2}.

\begin{deluxetable}{lccC}
\tablenum{1}
\tablewidth{0pt}
\tablecaption{Results of the quasi-simultaneous X-ray/optical observations. We provide the net 0.3--10.0~keV count rate for the X-ray observations and the V-band magnitudes for the optical ones.} 
\tablehead{\colhead{Date} & \colhead{Telescope} &
\colhead{$T_{\mbox{exp}}$, s}  & \colhead{Count rate or V magnitude}}
\startdata
2017-04-30 & Zeiss-1000 & 900 & $21.64\pm0.31$ \\
2017-05-02 & Swift/XRT & 2055 & $(5.0\pm 1.7) \times 10^{-3}$ \\ \hline
2017-07-21$^\dagger$ & BTA & 330 &	$21.43\pm0.16$ \\
2017-07-24$^*$ & Swift/XRT & 1770 &$<2.4\times 10^{-3}$ \\
2017-07-27$^\dagger$ & Zeiss-1000 & 2400 & $21.48\pm0.11$ \\ \hline
2017-08-15 & BTA & 360 & $20.86\pm0.12$ \\
2017-08-16 & Swift/XRT & 1690 & $(1.6 \pm 0.3)\times 10^{-2}$ \\ \hline
2018-04-11 & Zeiss-1000 & 1800 & $20.87\pm0.07$ \\
2018-04-11 & Swift/XRT & 1840 & $(2.9 \pm 0.4)\times 10^{-2}$ \\ \hline
2019-04-06 & Zeiss-1000 & 1200 & $21.57\pm0.09$ \\
2019-04-13$^*$ & Swift/XRT & 230 & $<9\times 10^{-3}$ \\
\enddata
\tablecomments{$^\dagger$This pair of the optical observations was averaged. $^*$Only a 90\% upper limit to the count rate is provided due to low statistics.}
\label{Tab1}
\end{deluxetable}

Starting from April 2017, we carried out five quasi-simultaneous observations with {\it Swift}/XRT and optical telescopes of SAO RAS in V band (the Zeiss-1000 equipped with a CCD photometer and the 6-m telescope BTA with the multi-mode focal reducer SCORPIO \citep{Afanasiev2005scorpio}). The time intervals between the X-ray and optical observations were within 1 day in the bright state of the ULX and in the range from 1.6 to $\simeq6$ days in the faint state (Table~\ref{Tab1}). An aperture photometry of the UGC\,6456~ULX counterpart was done with the APPHOT package in IRAF. To check the object for variability we used four stars with colors similar to those of the object as reference sources. The seeing was from $1\farcs0$ to $2\farcs1$ in all the observations. The obtained magnitudes were converted to the Vegamag system using three relatively bright isolated stars whose magnitudes were measured from the HST/WFPC2/F555W data.

During these quasi-simultaneous observations the {\it Swift} count rate changed by at least 6 times, and the visual magnitude by $\Delta$m$_V \approx 0.8$ mag (from M$_V \approx -6.8$ to M$_V \approx -7.6$).
We found a correlation between the fluxes in these two bands (Fig.~\ref{Fig2}, right panel) with a Pearson correlation coefficient of $0.9\pm 0.3$, which corresponds to the p-vales of 0.07. Moreover, since the observations are not purely synchronous, the observed correlation might be underestimated because the X-ray flux demonstrate rapid changes (up to 3 times within one day).

\section{Optical spectroscopy}
\label{sect:spec}

The optical spectra were obtained with the BTA/SCORPIO in 2015--2019; the 1\arcsec\ slit and four grisms were used. In our work we use one observation conducted with the VPHG1200B grism (spectral range is 3600-5400 \AA, resolution $\approx5.5$ \AA), seven observations with VPHG1200G (4000-5700 \AA, $\approx5.3$ \AA), one with VPHG1200R (5700-7500 \AA, $\approx5.3$ \AA) and the remaining four observations with VPHG550G (3500-7200 \AA, $\approx12$ \AA). Seeing varied from 1\arcsec\ to 2\arcsec. Data reduction was carried out with the LONG context in MIDAS using standard algorithm. The spectra were extracted with the SPEXTRA package \citep{Sarkisyan2017SPEXTRA}.

\begin{figure*}
\centering
\includegraphics[angle=0,scale=0.75]{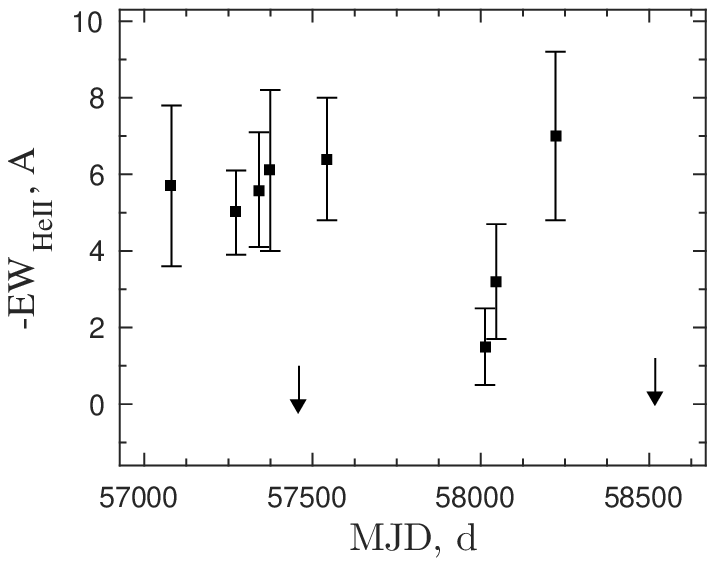}
\hspace{0.5cm}
\includegraphics[angle=0,scale=0.75]{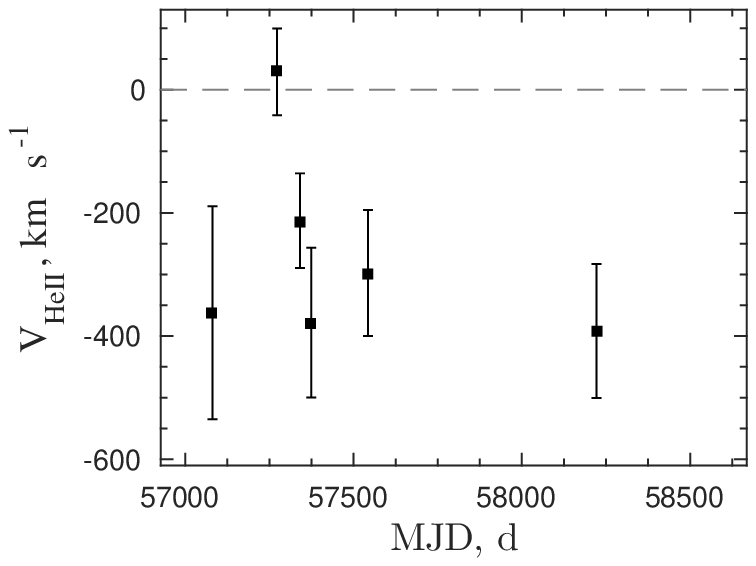}\\
\includegraphics[angle=0,scale=0.75]{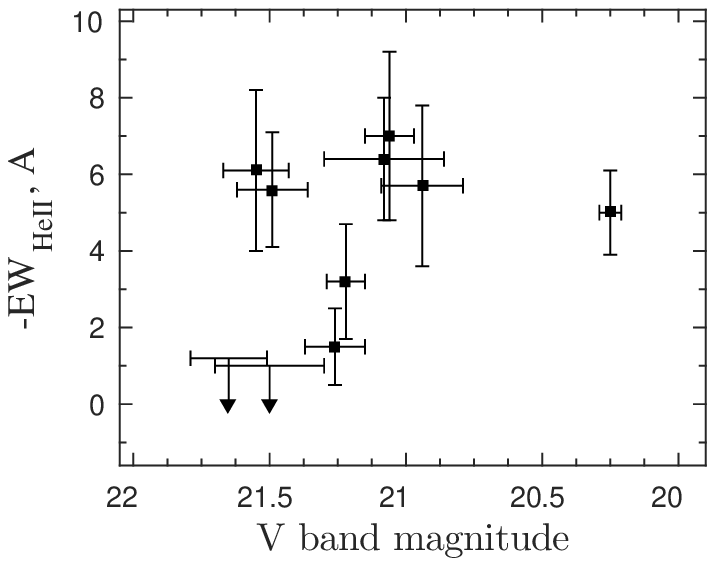}
\hspace{0.5cm}
\includegraphics[angle=0,scale=0.75]{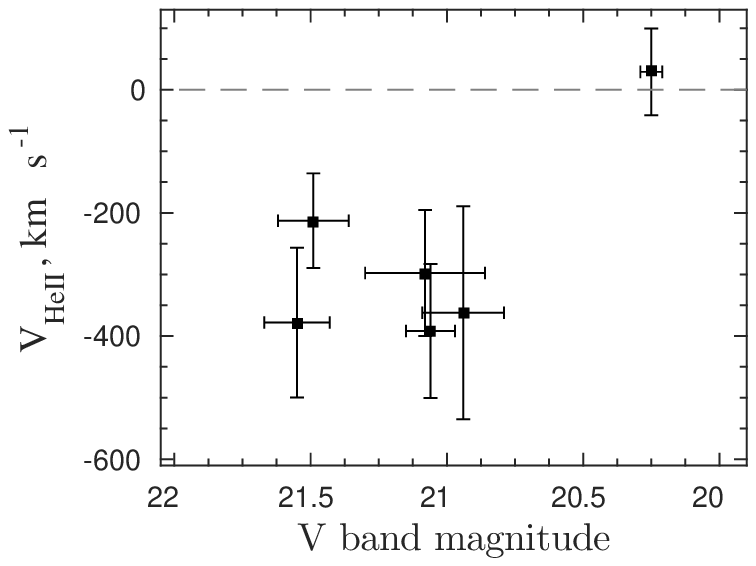}
\caption{{\it Left top panel}: variations of the equivalent width (EW) of the He\,II~$\lambda4686$ emission line. The arrows indicate an 90\%\ upper limit to the EW for those observations where the line was nearly undetectable. Two pairs of adjacent observations taken with a time interval of about one day with the VPHG1200G and VPHG550G grism were averaged. {\it Right top panel}: radial velocity curve measured from the He\,II~$\lambda4686$ emission line. Two data points near MJD=58000 are omitted due to a very low accuracy of the velocity measurements. {\it Left bottom panel}: the equivalent width (EW) of the He\,II line versus the V-band magnitude. {\it Right bottom panel}: the He\,II radial velocity versus the V-band magnitude of UGC\,6456~ULX.} 
\label{Fig3} 
\end{figure*}

Two spectra with the best signal-to-noise ratio are shown in the right panel of Fig.~\ref{Fig1}. The spectra are not simultaneous. The first one taken with the VPHG1200R grism was obtained on 2015 February 23, whereas the second with VPHG1200G on 2015 September 7. A broad He\,II~$\lambda4686$ line is clearly seen. Its width after correction for the spectral resolution is FWHM$=880\pm80$ km\,s$^{-1}$. The blue spectrum contains also a weak broad component of the H$_{\beta}$ emission line with the width comparable to that of the He\,II line and also possible contains H$_\gamma$ emission and H$_\delta$ absorption lines. In the red spectrum, there are relatively broad He\,I emission lines with FWHMs (after correction for the spectral resolution) of $260\pm80$ km\,s$^{-1}$ (He\,I~$\lambda5876$) and $290\pm50$ km\,s$^{-1}$ (He\,I~$\lambda7065$). We also detected He\,I~$\lambda6678$ line, but we can not reliable estimate its width due to low signal-to-noise ratio. A very strong and asymmetric H$_\alpha$ line has wings extending over 90 \AA.

Narrow emission lines ([O\,III]~$\lambda\lambda4959,5007$, [S\,II]~$\lambda\lambda6716,6731$ as well as the narrow components of the H$_{\beta}$ and H$_\alpha$) can be formed in a nebula near the source (which, however, may be too compact and too weak to be seen in Fig.1).

All the wide lines are highly variable. We have investigated a behavior of the He\,II~$\lambda4686$ line because it is not contaminated by the nebula emission. Its equivalent width (EW) vary from $\approx -7$ to $\gtrsim -1$~\AA\ (when the line is nearly undetectable, the left panels of Fig.~\ref{Fig3}) and the width from $\approx500$ to $\approx900$ km\,s$^{-1}$. Also we found changes in the radial velocity up to 400 km\,s$^{-1}$ (the right panels of Fig.~\ref{Fig3}) but additional observations are required to distinguish whether the nature of these changes is periodic or stochastic. Barycentric corrections are taken into account. We note that for the VPHG1200B and G observations the accuracy of the radial velocity measurements is better than 15 km\,s$^{-1}$ and better than 30 km\,s$^{-1}$ for the VPHG550G ones.

All the optical spectra were taken together with V-band images which allow us to check a possible correlation between the He\,II equivalent width and the optical brightness of the source. The photometry was performed by the same method as described in section~\ref{sect:XrayOptica}. We found that the object V magnitude varied during these observations by $\Delta$m$_V =1.40\pm0.15$ mag (from $21.65 \pm 0.14$ to $20.25 \pm 0.04$). The equivalent widths and radial velocities of the He\,II~$\lambda4686$ line versus the obtained V magnitudes are shown in bottom panel of Fig.~\ref{Fig3}.
It is seen that, in general, the $|\mbox{EW}|$ is held near 5--6~\AA\ regardless of the source optical brightness, however, when the source is faint the line may sometimes weaken. The constancy of the line EW implies that the total flux in the line increases as the source become brighter. 
\cite{Binder2018} analysing simultaneous optical spectroscopy--X-ray observation of NGC\,300~ULX-1 found that the X-ray flux and the flux in the He\,II~$\lambda4686$ are correlated. Our spectroscopic and X-ray observations are not simultaneous, nevertheless, combining the relation between the line flux and m$_V$ with the relation between m$_V$ and the X-ray count rate (Fig.~\ref{Fig2}), we can conclude that similar correlation may occur in the case of UGC\,6456~ULX as well. However new observations are still needed to prove this more reliable.
Another interesting feature is that the He\,II line showed the largest jump of the radial velocity when the source was in the brightest state.

Using observed ratios of the hydrogen lines in the nebula (H$_{\gamma}$/H$_{\beta}$ and H$_\alpha$/H$_{\beta}$), we determined the reddening value of A$_V = 0.2\pm0.1$ mag. In our calculation we assumed Case B of photoionization \citep{Osterbrock2006book}. However, we have to note that it is unknown whether the nebula is associated with the ULX or just lie on the line of sight. To clarify this, one needs additional high resolution observations in narrow filters. Nevertheless, all the nebulae in the area of the object show approximately the same reddening, therefore, we believe that the estimate of A$_V = 0.2\pm0.1$ mag is reliable regardless of the nature of the nebula whose lines are observed in the spectrum.

\section{Conclusions}
We have identified UGC\,6456~ULX with the single optical source that falls into the astrometry error circle. Discovered optical counterpart is one of the brightest among all identified ULXs (M$_V$ up to $-8.24 \pm 0.11$; the A$_V$ uncertainty was taking into account). The object demonstrates high optical and X-ray variability, which has an amplitude similar to that observed in the well studied ULX with a neutron star NGC\,7793~P13 \citep{Furst2018P13orb}. We found a correlation between the optical and X-ray fluxes, which may indicate that the optical emission originates from re-processing of X-rays in the outer parts of the strong wind from the supercritical disk. The existence of such a wind is confirmed by the broad hydrogen and helium emission lines in the optical spectra, that are typical for many spectroscopically studied ULXs \citep{Fabrika2015}. 

Nevertheless, some ULXs show optical spectra that differ to varying degree from each other and from the spectrum of UGC\,6456~ULX. For example, M101~ULX-1 shows only broad helium and nitrogen emission lines belonging to a Wolf-Rayet donor \citep{Liu2013}, which indicates that this system is more evolved than UGC\,6456~ULX, where the donor must be a hydrogen rich star. Optical spectra of NGC\,7793~P13 are dominated by emission from a B9Ia donor star \citep{Motch2014}: the spectra exhibit strong high-order Balmer absorption lines, as well as relative weak absorption lines of He\,I, Si\,II, Fe\,II, and Mg\,II. More typical for ULXs emission lines of hydrogen (up to H$_{\gamma}$), He\,II~$\lambda4686$ and Bowen blend CIII-NIII are also present in the source spectrum; these lines may originate either from some structures of the accretion disk (like a disk bulge at the point where the stream of matter escaping L1 impacts the edge of the disc, \citealt{Motch2011}) or from the accretion disk wind. In contrast to this, the ultraluminous pulsar NGC\,300~ULX-1 have a much more complex emission spectrum \citep{Binder2018,Heida2019} indicating the presence of a $\approx1000$ km\,s$^{-1}$ outflow and ionization by X-rays. In particular, the He\,II~$\lambda4686$ line luminosity was found to be correlated with the soft X-ray emission. Our results hints that in the case of UGC\,6456~ULX similar correlation also may be present.

We believe that uniquely high optical brightness of UGC\,6456~ULX, isolation from other bright stars together with high variability in both X-ray and optical bands make this source the best target for studying the nature of optical emission from ULXs.

\normalem
\begin{acknowledgements}
The study was funded by RFBR according to the research project 18-32-20214. This work made use of data supplied by the UK Swift Science Data Centre at the University of Leicester. Also we are grateful to A. Moskvitin and O. Spiridonova for their observations on Zeiss-1000. Observations with the SAO RAS telescopes are supported by the Ministry of Science and Higher Education of the Russian Federation (including agreement No05.619.21.0016, project ID RFMEFI61919X0016).

\end{acknowledgements}
  


\end{document}